\documentclass[12pt,preprint]{aastex}

\begin{document}

\title{Effects of Gravitational Microlensing on P-Cygni Profiles of Type Ia Supernovae}

\author{Hamed Bagherpour %
\footnote{hamed@nhn.ou.edu %
} , David Branch %
\footnote{branch@nhn.ou.edu %
} , Ronald Kantowski %
\footnote{kantowski@nhn.ou.edu %
} }

\affil{University of Oklahoma, Department of Physics and Astronomy,\\
 Norman, OK 73019, USA }

\begin{abstract}
A brief description of the deformed spectra of microlensed SNe Ia
is presented. We show that microlensing amplification can have significant
effects on line profiles. The resonance-scattering code
SYNOW is used to compute the intensity profile in the rest frame
of the supernova. The observed (microlensed) spectral lines are predicted
assuming a simple stellar-size deflector, and are compared
to unlensed cases to show the effects microlensing by solar-size deflectors can have
on spectral lines. We limit our work to spherically symmetric deflectors. 
\end{abstract}


\section{Introduction}

It is known that amplification of a light source due to microlensing
can affect the spectral line profiles if different parts of the
profile originate from different emitting regions of the source (e.g., if the emitting
region has a ring-like structure) and if the sizes of these regions are comparable to the 
characteristic lensing radius \citep{KayRefSta86}.
Evidence of such effects on line profiles of broad-absorption-line (BAL) quasars was suggested 
by spectroscopic observations of the multiple images of the strongly lensed quasar
H1413+117 \citep{Ang90}. Spectral differences observed between
the four images of this quasar were investigated by \citet{Hut93} 
using the lens model of \citet{ChaRef84}.

One expects these `chromatic amplifications' to be observed in a microlensed
type Ia supernova as well. SNe Ia are well-studied extended light sources consisting 
of a central continuum source surrounded by a rapidly
expanding atmosphere. The atmosphere accounts for the formation of the observed
P-Cygni profiles in the spectral lines of these objects. 
In $\S$ 2 we show how microlensing by a simple (point-mass) Schwarzschild deflector 
can deform these P-Cygni line profiles. 
Results of lensing a source at $z_{s}=1$ by a deflector at $z_{d}=0.05$ are 
presented in $\S$ 
3. A flat Friedmann-Lama\^{\i}tre-Robertson-Walker cosmological model
with $\Omega _{m}=$ 0.3, $\Omega _{\Lambda }=$ 0.7, and h$_{100}=$ 0.67  
is assumed to calculate source and deflector distances.

\section{Microlensing of a Supernova as an Extended Source}

\subsection{Line Profiles of Type Ia Supernovae}

P-Cygni profiles are characterized by emission lines together
with corresponding blueshifted absorption lines. The latter is produced
by material moving away from the source with either relativistic
velocities \citep{HutSer90} or nonrelativistic velocities
\citep{Bea29}. For an explosive expansion the material in each plane
perpendicular to the line of sight has a fixed component of velocity $V_{r}$ 
toward the observer (Fig. 1). 

Using the normalized,
projected radius $P$ (i.e., the photosphere at $P=1$, see Fig. 1), the unlensed flux is 
defined as
\begin{equation}
F_{\lambda }=\int _{0}^{2\pi }\int _{0}^{P_{max}}I_{\lambda }(P,\theta )\, P\, dP\, d\theta \: ,
\end{equation}
where $P_{max}$ is the normalized, projected radius of the supernova and
the plane polar angle $\theta $ is measured in the projected source plane. 

To compute the intensity $I_{\lambda }$, we used a resonance-scattering
synthetic spectrum produced with the fast, parameterized supernova
synthetic-spectrum SYNOW. This
code is often used for making and studying line identifications and initial coarse
analysis of the spectra during the photospheric phase of the supernova \citep{Bra03, Bra05}. 
The code assumes a spherically symmetric and sharp photosphere that 
emits a blackbody continuum characterized by temperature $T_{bb}$. 

In SYNOW, the expansion velocity is proportional to radius (homologous
expansion: $v=r/t$), as expected for matter coasting at a fixed velocity after
an impulsive ejection with a range of velocities. Line formation is
treated in the Sobolev approximation \citep{Sob58} and occurs by
resonance scattering of photons originating from the photosphere. Line blending
is treated in a precise way, within the context of the Sobolev approximation.
Each line optical depth $\tau $ is taken to decrease exponentially with radius 
[See \citet{JefBra90} and \citet{Fis00} for more details].
The code does not calculate ionization ratios or rate equations; it
takes line identifications to estimate the velocity at the photosphere and
the velocity interval within which ions are detected. These quanities give
constraints on the composition structure of the ejected matter.
 
SYNOW calculates the intensity emitted from each zone (concentric
annuli) of the projected source. These intensity profiles show the
absorption features for various $P<1$ as well as emission features for
$P>1$. The weighted sum of these intensities over the projected surface of 
the supernova (eq. [1]) gives the synthetic flux profile. 
Figure 2 shows the calculated intensity profiles of sodium. These profiles 
are obtained for optical depths of $\tau = 1$ (upper panel) and $\tau = 1,000$ 
(lower panel). The intensities are calculated for 
$P>1$, $P=0$, and $P$ just below 1. The asymmetry seen in the emission
features is caused by applying the relativistic Sobolev method \citep{Jef93} 
in the SYNOW code which, in turn, introduces a Doppler boosting in the profiles.
Note that to relate the quantities in the comoving and observer frame, one
has to use the transformation
\begin{equation}
I_{\lambda}=I_{\lambda_{o}} \, \frac{\lambda^{5}_{o}}{\lambda^{5}}\: ,
\end{equation}
where the quantities with subscript `o' denote those
measured in the comoving frame \citep{Mih78}.

\subsection{Differential Amplification}

Microlensing of an extended source such as a type Ia supernova by an isolated compact 
mass results in two images, both in line with the source and deflector projected on
the sky plane. The `primary' image, $r_{p}$, lies on the same side
of the deflector as the source while the `secondary' image, $r_{s}$, is on the
opposite side. For solar-mass size deflector the angular separation of the two images is
of the order of micro arcseconds and as a result, they are seen as
a single object as expected
for microlensing events \citep{SchEhlFal92}. The
apparent brightness of this `single' image differs from that of the
unlensed source and is proportional to the apparent area of the image,
meaning that the brightness of a source is amplified by a factor 
\begin{equation}
Amp=\frac{A_{o}}{A}\: ,
\end{equation}
where $A_{o}$ is the area of the image and $A$ is the area of the
source both projected on the sky plane. See \citet{Bag05} for details. 

If the different parts of the spectral profile come from different
parts of the source (which is the case for the concentric annuli of
an isotropically expanding type Ia supernova), and if the emitting regions are not much bigger
than the characteristic lensing radius \citep{KayRefSta86}, one may expect to
see not only a rescaling in the observed flux, but also a 
deformation in the line profiles. For a lensed supernova, the observed
line profile becomes 
\begin{equation}
F_{\lambda }=\int _{0}^{2\pi }\int _{0}^{P_{max}}I_{\lambda }(P,\theta )\, Amp(P,\theta )\, P\, dP\, d\theta \: ,
\end{equation}
where $Amp(P,\theta )$ is the amplification of the surface element
centered at point $(P,\theta )$. Assuming that a supernova explodes isotropically, 
the altered flux becomes 
\begin{equation}
F_{\lambda }=\int _{0}^{P_{max}}I_{\lambda }(P)\, Amp(P)\, P\, dP\: ,
\end{equation}
where
\begin{equation}
Amp(P)\equiv \int _{0}^{2\pi }Amp(P,\theta )\, d\theta \: .
\end{equation}
For an annulus with the width $\delta P$ bounded
by inner and outer radii of $P_{-}\equiv P-\frac{\delta P}{2}$ and
$P_{+}\equiv P+\frac{\delta P}{2}$, the amplification becomes
\begin{equation}
Amp(P)=\frac{S_{+}-S_{-}}{\pi \left(P_{+}^{2}-P_{-}^{2}\right)}\: ,
\end{equation}
where 
\begin{equation}
S_{\pm }=\int _{-\frac{\pi }{2}}^{\frac{\pi }{2}}P_{\pm }\left(P_{\pm }+l\, \sin \varphi \right)\sqrt{1+\frac{4r_{E}^{2}}{l^{2}+P_{\pm }^{2}+2P_{\pm }l\, \sin \varphi }}d\varphi \: .
\end{equation}
(See \citet{Bag05}). In the above equation, $l$ is the distance between
the deflector and supernova, and $r_{E}$ is the radius of Einstein ring. All 
distances are projected on the deflector plane. 

The amplification can be calculated using elliptical integrals \citep{WitMao94, MaoWit98}.
Figure 3 shows amplification as a function of the expansion velocity 
of the projected annuli for several values of
$l$ in terms of the Einstein ring $r_{E}$. Notice that each curve peaks
at the velocity corresponding to the annulus which is partially obscured
by the deflector.

\section{Lensed Profiles}

In this section we calculate the deformed spectral profiles of microlensed
SNe Ia. To calculate the differential amplification, we need to specify
the distance $l$ between the supernova and the deflector projected
on the plane of the deflector (sky plane), normalized by the Einstein
ring radius $r_{E}$,
\begin{equation}
u_{A}\equiv \frac{l}{r_{E}}\, .
\end{equation}
The radius $r_{E}$ is determined by the mass of the deflector $m_{d}$ and the weighted
distance (of luminosity distances) $D$,
\begin{equation}
r_{E}=\sqrt{\frac{4Gm_{d}D}{c^{2}}}\, ,
\end{equation} 
where $D \equiv D_{ds}D_{d}/D_{s}$ in which, $D_{s}$, $D_{d}$, and $D_{ds}$ are 
the respective observer-source, observer-deflector, and deflector-source 
distances. These are angular size distances calculated
adopting a $(\Omega _{m},\, \Omega _{\Lambda },\, h)=(0.3,\, 0.7,\, 0.67)$
cosmology. We assume that
the source is at redshift $z_{s}=1.0$ and is lensed by a deflector
located at redshift $z_{d}=0.05$. The point-like deflector has a
mass of 1 M$_{\odot }$ and when positioned at different projected distances
it magnifies parts of the extended source differently. The Einstein radius $r_{E}$ for
the above deflector mass and distances is $\sim1301$ AU.

We have considered an SN Ia with fixed radius of 178 AU, corresponding
to that of a supernova with a maximum atmospheric speed of 30,000
km s$^{-1}$ at eighteen days after the explosion. The radius of the supernova 
projected on the deflector plane, $r_{SN}$, is $\sim20$ AU, with a photospheric radius of
$r_{Ph}\approx8$ AU. The black body continuum temperature $T_{bb}$ is taken to be 14,000 degrees. 
Optical depths $\tau $ are taken to vary exponentially for each line as 
\begin{equation}
\tau (v)=\tau _{o}\, exp\left(-\frac{v}{v_{e}}\right)\, ,
\end{equation}
where $v$ is the expansion velocity of each layer and $v_{e}$ is
the corresponding e-folding velocity (e.g., 1,000 km s$^{-1}$ for sodium).

\setcounter{footnote}{0}

We used SYNOW to calculate the unlensed
sodium lines as well as their corresponding lensed profile for different
optical depths $\tau _{o}$ and normalized distances $u_{A}$ to show
the effect microlensing can have on a single, clean line. Figures
4 through 7 show the results of such calculations for $\tau _{o}=1$ and $1,000$,
and $u_{A}=0$ and $1/128\ (\approx0.008)$\footnote{We changed $u_{A}$ in terms of 
negative powers of 2 to let it go to zero assymptotically.}.
In these diagrams, flux is plotted in
an arbitrary unit as a function of wavelength. As can be seen in Figure 4, the
emission feature is reduced with respect to the absorption feature because the $P<1$
region is magnified much more than the $P>1$ region. The narrowing of the absorption
dip as well as an overall shift of the lensed curve to the left is due to the 
extreme amplification of the $P=1$ annulus. This is the result of the source-deflector alignment 
(Fig. 3) which causes the area with
the highest blueshift ($P=1$) to get the highest amplification. With $u_{A}=1/128$ (Fig. 5),
the emission feature of the apparent line profile is magnified while the absorption
feature does not change remarkably because the amplification curve maximizes 
outside $P<1$ region and flattens inside (Fig. 3). Figure 6 is the same as
Figure 4 with $\tau = 1,000$. In this figure, we once again notice the effect of extreme 
amplification of the central zone of the source in the form of a slight shift 
of the apparent curve toward lower wavelengths. Contrary to
Figure 4, we do not encounter sharp dips here because each dip in the (unlensed) intensity
curve is 7,000 km s$^{-1}$ wide. As expected, moving 
the deflector away from the line of sight to the source results in a stronger emission
component while the absorption dip does not vary remarkably (Fig. 7).    
 
Figures 8 and 9 show the same calculations as those of Figures 4 to 7 for a 
SYNOW synthetic spectrum that resembles 
that of a SN Ia near maximum light with $u_{A}=0$ and $1/128$, respectively.
We have normalized the lensed profile at $\lambda = 7,000$ \AA.
Because noticeable deformation of the profiles
appear only when the deflector is almost aligned with the source (small
values of $u_{A}$), we did not include results for $u_{A}>1/128$ (i.e., $l > 10.17$ AU). 
With $u_{A}=0$ (Fig. 8), central parts of supernova are amplified more than the rest which,
as explained for Figures 4 and 6, results in a slight shift toward lower wavelengths. 
Again, emission features are demagnified
with respect to the absorption components. The observed spectral lines show sharp dips 
because the input value of optical depth $\tau$ for each line is not too high. Figure 9
is the same as Figure 8 but with $u_{A}=1/128$. Here, the light coming from the $P>1$ 
area carrying emission features has higher amplification compared to the blueshifted 
light emitted from $P<1$ and, as expected, the emission component of the P-Cygni 
features is magnified more than the absorption part.  

The change in the profiles can, in general, be summarized as a net increase or decrease of
the absorption component relative to the emission one. The apparent change in either 
component may be so strong that an
emission feature could look like a typical P-Cygni profile. To see this effect, the
projected source (supernova) must be very close to the deflector on the sky plane 
($l \ll r_{E}$). Larger impact parameters produce less dramatic deviations from the unlensed 
profile and could easily be attributed to the intrinsic diversity in spectra of type Ia 
supernovae rather than gravitational lensing. 
The deflector redshift used here ($z_{d}=0.05$) is not the most likely redshift for
microlensing but results in a remarkable amplification
gradient necessary for noticeable deformation of P-Cygni profiles.
In general, the probability of microlensing cosmologically distant light source
is not negligible, and can exceed 1\% for a source located at $z_{s}=1$ and beyond 
\citep{Mye95, WyiTur02, ZakPopJon04}. 
However, for the small values of $u_{A}$ used here
the probability of observing such deformations drops below 0.001\% \citep{Bag05}.

It should be noted that we have ignored any contribution to lensing from nearby stars and the
deflector's parent galaxy. For most microlensing events the amplification due to the host galaxy 
is expected to introduce a 
small amplification gradient across the supernova which merely rescales the line profile 
without introducing any significant deformation effects. Macrolensing by the galaxy should 
mainly bias the supernovae detection. The rescaling effect is 
compensated for by normalizing the lensed profile in order to compare the apparent and unlensed
profiles, as done in Figures 4 through 9. When large amplification gradients are
introduced, either by the galaxy or by microlensing stars nearby, a more complicated lens models will 
be required.

\section{Conclusion} 

We have shown that microlensing can significantly affect the P-Cygni
profile of a cosmologically distant type Ia supernova. We restricted
our calculation to $z_{s}=1$ and $z_{d}=0.05$ in the commonly used
$\Omega _{m}=0.3$, $\Omega _{\Lambda }=0.7$ flat cosmological model
with h$_{100}=0.67$. We found that microlensing can not only increase the
flux magnitude but also can cause a change in its line profiles.
Microlensing can cause the features in the spectral lines to be blueshifted
with respect to the original spectrum and in general, results in a net 
increase or decrease of the absorption component relative to the emission 
component.

We calculated the deformed line profiles for special cases where the deflector
is extremely close to the line of sight to the source. 
Due to the low probability of microlensing events occuring with such small values of $u_{A}$,
a large population of supernovae (around $10^{5}$) would have to be surveyed to observe a
single case of deformation in P-Cygni profiles of type Ia SNe. Also, large
deformations demand using more complicated lensing models.

\acknowledgements{}

H. B. wishes to thank Darrin Casebeer for helpful discussions. This work
was in part supported by NSF grant AST0204771 and NASA grant NNG04GD36G.

\clearpage

\figcaption{}
Schematic view of a type Ia supernova as seen
by an observer. The normalized radius $P$ and planar speed $V_{r}$ are shown. 
\label{f1}

\figcaption{}
Synthetic intensity profile of a type Ia supernova,
showing absorption and emission features for $P<1$, $P>1$, and P just below 1
for $\tau=1$ (upper panel) and $\tau=1,000$ (lower panel). The emission feature in 
the upper panel is scaled up.
\label{f2}

\figcaption{}
This figure shows the amplification curves of any point on the projected source 
as a function of expansion velocity for different values of $l$. The value
of photospheric expansion speed determines the zone (absorption or emission)
with higher amplification. 
\label{f3}

\figcaption{}
Amplified line profile of sodium for $\tau =1$ and $u_{A}=0$. The deflector has a
mass of 1M$_{\odot}$. 
\label{f4}

\figcaption{}
Same as Fig. 4, with $\tau =1$ and $u_{A}=1/128$.
\label{f5}

\figcaption{}
Same as Fig. 4, with $\tau =1,000$ and $u_{A}=0$.
\label{f6}

\figcaption{}
Same as Fig. 4, with $\tau =1,000$ and $u_{A}=1/128$. 
\label{f7}

\figcaption{}
Amplified spectral lines of a SYNOW spectrum that resembles the maximum-light
spectrum of a SN Ia, with $u_{A}=0$ and m$_{d}=1$M$_{\odot}$
\label{figure8-9}

\figcaption{}
Same as Fig. 8, with $u_{A}=1/128$ . 
\label{f9}

\clearpage

\plotone{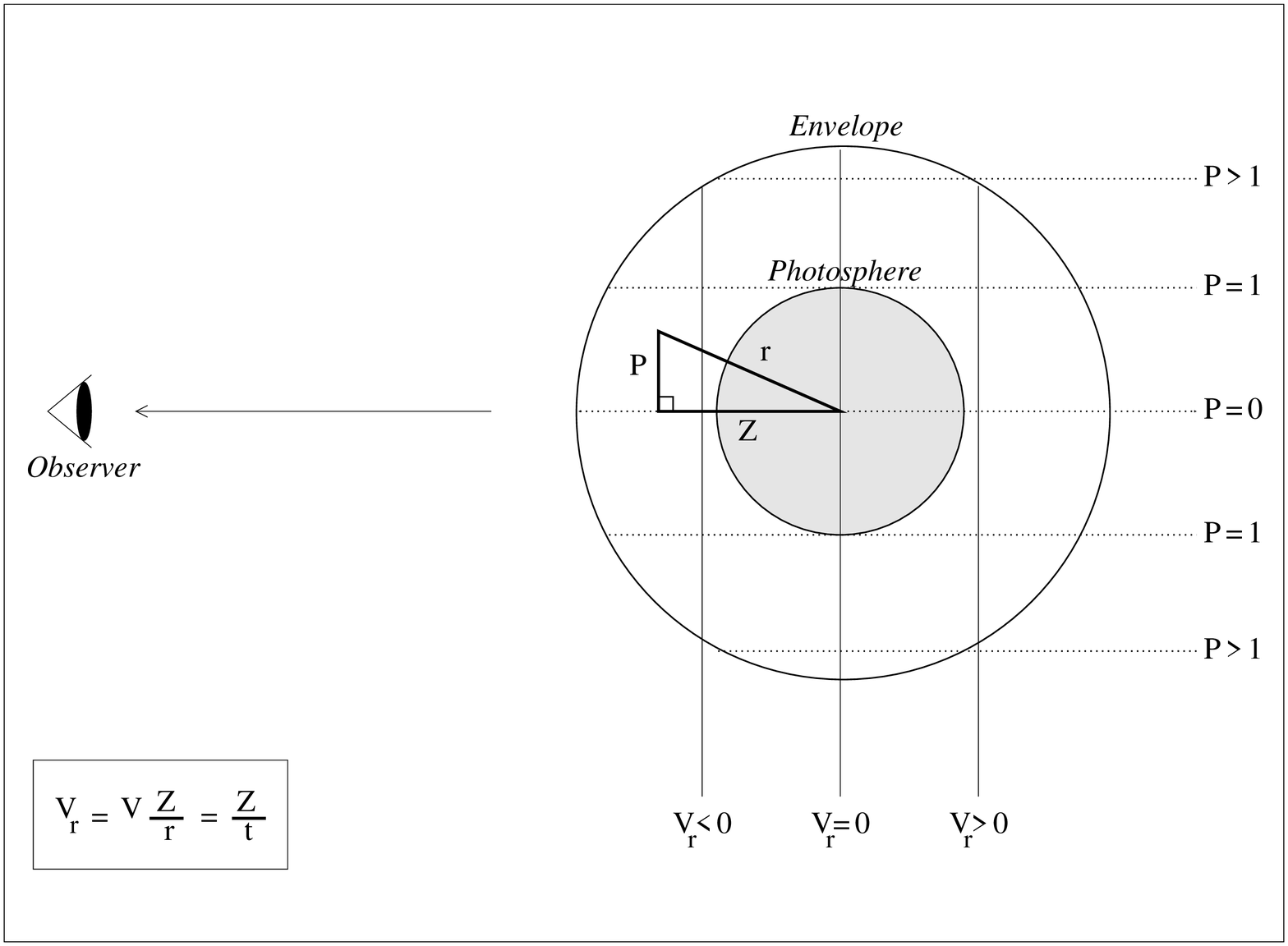}
\begin{center} Figure 1 \end{center}
\eject

\epsscale{0.65}
\plotone{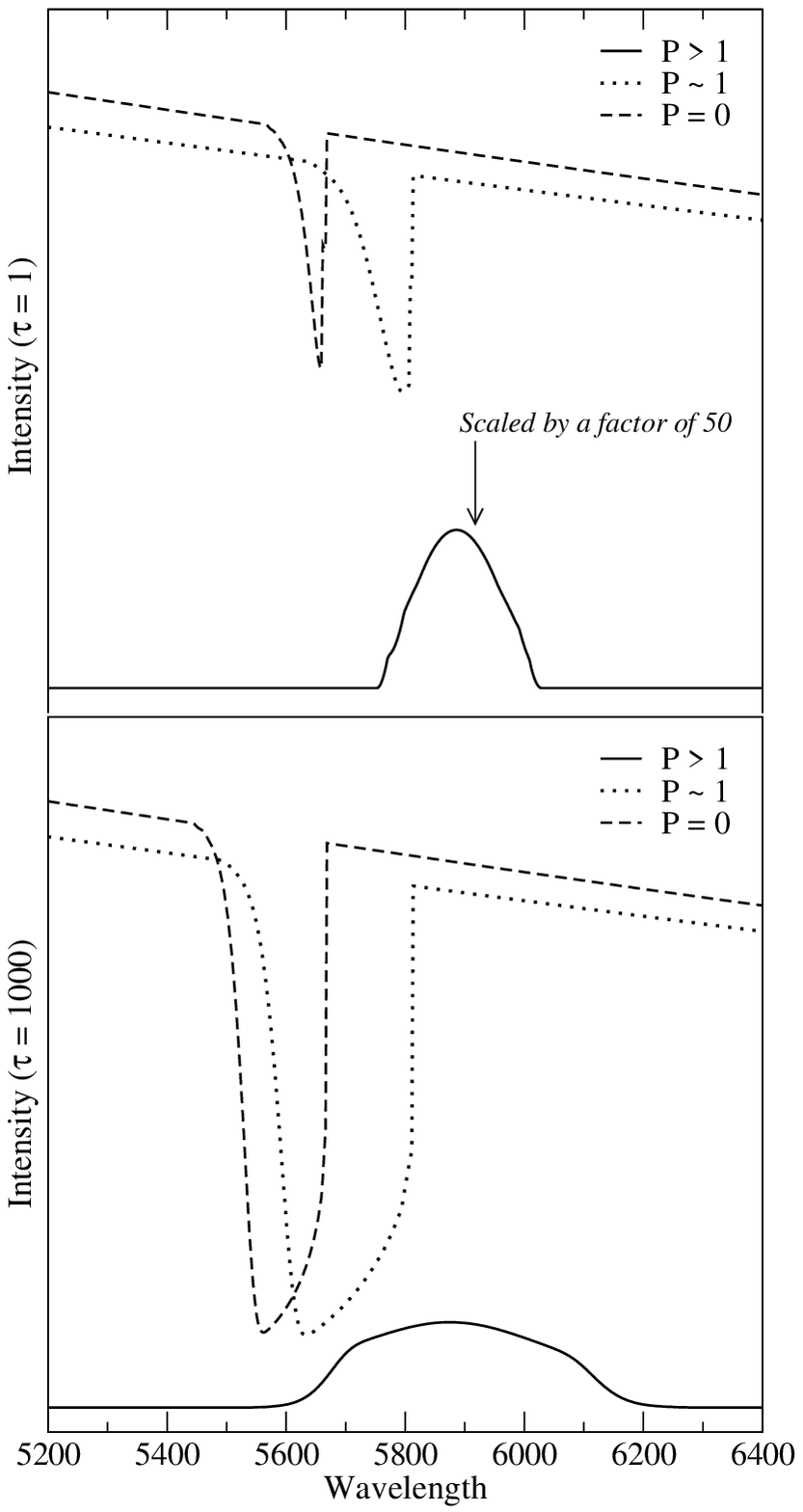}
\begin{center} Figure 2 \end{center} 
\eject

\epsscale{1}
\plotone{f3.eps}
\begin{center} Figure 3 \end{center}
\eject

\epsscale{0.65}
\plotone{f4.eps}
\begin{center} Figure 4 \end{center}
\eject

\epsscale{0.65}
\plotone{f5.eps}
\begin{center} Figure 5 \end{center}
\eject

\epsscale{0.65}
\plotone{f6.eps}
\begin{center} Figure 6 \end{center}
\eject

\epsscale{0.65}
\plotone{f7.eps}
\begin{center} Figure 7 \end{center}
\eject

\epsscale{1}
\plotone{f8.eps}
\begin{center} Figure 8 \end{center}
\eject

\plotone{f9.eps}
\begin{center} Figure 9 \end{center}
\eject

\end{document}